\documentstyle[12pt]{article}
\begin{document}
\newcommand{\scr}{\sin^2 \hat{\theta}_W (m_Z)}
\newcommand{\sef}{\sin^2 \theta_{eff}^{lept}}
\newcommand{\msbar}{\rm{\overline{MS}}}
\newcommand{\be}{\begin{eqnarray}}
\newcommand{\en}{\end{eqnarray}}
\newcommand{\kcar}{\hat{k}}
\newcommand{\scar}{\hat{s}}
\newcommand{\cc}{\hat{c}}
\newcommand{\amz}{\alpha(m_Z)}
\newcommand{\amzc}{\hat{\alpha}(m_Z)}
\newcommand{\arun}{\alpha_{\rm run}}
\newcommand{\alc}{\hat{\alpha}}
\newcommand{\nn}{\noindent}
\newcommand{\ewc}{electroweak corrections\ }
\newcommand{\dres}{(\Delta r)_{\rm res}}

\nn     \hspace*{10.5cm} NYU--TH--93/11/01 \\
\hspace*{11.2cm} November 1993 \\

\vspace*{2cm}

\centerline{\large{\bf How Strong is the   Evidence for Electroweak}}

\centerline{\large{\bf Corrections Beyond the Running of $\alpha$? }}

\vspace*{1cm}

\centerline{\sc Alberto Sirlin.}

\vspace*{1cm}

\centerline{ Department of Physics, New York University, 4 Washington
Place,}
\centerline{ New York, NY 10003, USA.}

\vspace*{1cm}

\begin{center}
\parbox{14cm}
{\begin{center} ABSTRACT \end{center}
\vspace*{0.2cm}
The evidence for electroweak radiative corrections not contained in
$\amz$ is examined.
At low energies there is very strong evidence in the analysis of the
unitarity of the CKM matrix.
At LEP and collider experiments the current direct signals are not  strong,
but there is substantial indirect evidence, which will likely
become sharper when $m_t$ is determined. The advantage of using
 $(\Delta r)_{res}$ as a measure of these effects is emphasized. In order to
improve the direct evidence, more accurate measurements of $m_W$ and the
on--resonance asymmetries are indicated.
}

\end{center}
\vfill
\newpage
\nn
The presence in electroweak physics of large corrections associated with the
running of $\alpha$ at the vector boson scale was emphasized long ago
\cite{1}. The detailed analysis of electroweak corrections not described
by the running
of $\alpha$ is also a matter of long standing among particle physicists
[2--5].
Recent discussions of signals for the latter effects have focussed on LEP
and collider studies \cite{6,7}. On the basis of the directly measured
values of $m_W$ and $m_Z$, Hioki concluded that there is some evidence
for their existence, but only at the 1$\sigma$ level \cite{6}.
Very recently, Novikov, Okun and Vysotsky (N--O--V) pointed out that a
Born approximation treatment based on $\amz$ and a suitable definition of
the weak mixing angle reproduces very well the most precise LEP
and collider information \cite{7}. From this observation they concluded
that there is no evidence for corrections beyond those associated with
$\amz$.
In their formulation, sharp constaints on $m_t$ emerge, as in many previous
analyses of electroweak data.
However, they are interpreted as stemming from the cancellation of top quark
effects against those of other virtual particles.

The aim of this paper is to examine the evidence for \ewc not described by
 the running of $\alpha$ at the $m_Z$ scale. For reasons that will become
clear later, we generically denote this parameter as $\alpha_{\rm run}$.
We first consider precision experiments at very low energies and then turn
our attention to LEP, SLC and collider physics.

Very strong evidence for \ewc not involving $\arun $ is found in the
analysis
of the universality of the weak interactions \cite{2}. It is well known
that the Standard Model (SM) leads to the relation
$|V_{ud}|^2+|V_{us}|^2+|V_{ub}|^2=1.$ This is simply
a consequence of the unitarity of the CKM matrix and indeed constitutes one
of the most fundamental predictions of the theory. We recall that
 $|V_{ud}|^2 $
is determined from the ratio of transition rates of superallowed Fermi
transitions and $\mu$ decay, while $V_{us}$ and $V_{ub}$ involve
consideration
of $\Delta S=1$ and $B$ decays. As an example, when $^{14} O$ is employed
and appropriate electroweak and nuclear overlap corrections are applied,
one currently finds \cite{8} $V_{ud}=0.9745\pm0.0005\pm0.0004$.
The first error includes statistical and nuclear overlap uncertainties,
while the second stems from the radiative corrections. In conjunction with
$V_{us}$ and $V_{ub}$ this leads to \cite{8}
\be
|V_{ud}|^2 +|V_{us}|^2+|V_{ub}|^2=0.9983\pm 0.0015  \ \ \ (^{14} O),
\label{ckm}\en
which is in good agreement with the theoretical expectations.
It is important to note, however, that the \ewc applied in obtaining
$|V_{ud}|^2$ are very large, namely 4.1\% in the case of $^{14} O$.
As a consequence, if these corrections and corresponding effects in
$|V_{us}|^2
$ are not included, Eq.(\ref{ckm}) becomes
\be
|V_{ud}|^2 +|V_{us}|^2+|V_{ub}|^2=1.0386\pm 0.0013  ,
\label{ckm2}\en
which differs from unity by $\approx 30$ times the estimated error.
 If the average of
the eight accurately measured superallowed Fermi transitions is employed
rather than $^{14} O$, there are small changes at the 0.2\% level in
Eq.(\ref{ckm}), but the sharp disagreement illustrated in Eq.(\ref{ckm2})
still holds. Thus, we see that the large \ewc obtained in the SM are
actually crucial for its survival as a viable theory. Because these
effects cannot be
 computed consistently in the local Fermi theory, where they are divergent,
their existence and the close agreement that follows from their
application can
 be regarded as a significant success of the SM.
We also note that $\beta$, $\mu$ and semileptonic decays occur at
$q^2\approx 0$ and that these large corrections are not associated
with the running of
 $\alpha$.

In order to discuss the evidence at LEP and collider physics one must first
precisely state how $\arun$ is defined. This concept is scheme--dependent
and, in fact, there are two frequently employed definitions,
\be
\amz\ =\ \alpha/{[1+e^2 {\rm Re}\Pi_{\gamma\gamma}^r(m_Z^2)]},
\label{amz}\en
and
\be
\amzc\ = \ \alpha/{[1-e^2 \Pi_{\gamma\gamma}(0)|_{\msbar}+...]}.
\label{amzc}
\en
Here $\Pi_{\gamma\gamma}(q^2)$ is the unrenormalized vacuum polarization
function and the ellipses in Eq.(\ref{amzc}) stand for additional
contributions involving $W$'s. The superscript $r$ in Eq.(\ref{amz})
indicates the conventional QED renormalization, namely the subtraction of
$\Pi_{\gamma\gamma}(0)$, while $\msbar$ in Eq.(\ref{amzc}) denotes the
$\msbar$
renormalization. In the latter case one subtracts the poles and associated
constants in dimensional regularization and chooses the 't--Hooft mass scale
$\mu$ to be equal to $m_Z$. Both definitions absorb the large logarithms
associated with the running of $\alpha$, but there is a 0.8\% numerical
difference between them. Specifically,
$\amz=(128.87\pm 0.12)^{-1}$ \cite{9} and $\amzc=(127.9\pm0.1)^{-1}$
\cite{10}.

Employing Eq.(\ref{amz}) and defining
\be
c_0^2 s_0^2 = \pi\amz/[\sqrt{2} G_\mu m_Z^2],
\label{sin}\en
so that $s_0^2=0.23118(31)$, N--O--V found that, at the time of their
writing,
the data for $m_W/m_Z$, $(g_A)_\ell$, $(g_V/g_A)_\ell$, $\Gamma_\ell$,
$\Gamma_h$ and $\Gamma_Z$ were reproduced within their 1$\sigma$
accuracies by a simple Born approximation (B. A.) calculation \cite{7}.
Here $(g_A)_\ell$ and $(g_V)_\ell$ are the effective couplings of $Z^0$
to leptons at $q^2=m_Z^2$, while
 $\Gamma_\ell$, $\Gamma_h$ and $\Gamma_Z$ represent the leptonic, hadronic
and total widths of $Z^0$. In particular, the predictions for $\Gamma_\ell$,
 $\Gamma_h$ and $\Gamma_Z$ are remarkably accurate and the corresponding
experimental values quite stable. On the other hand, $(g_V/g_A)_\ell$ is
very
sensitive to small variations in the electroweak data. For instance, a more
recent analysis which includes the preliminary high statistics
1992 LEP run, leads to $\sef=0.2321\pm0.0006$, as determined from the
on--resonance asymmetries \cite{11}.
This translates into $(g_V/g_A)_\ell=1-4\sef=0.0716(24)$, which differs from
 the B.A. prediction by 1.4$\sigma$. Also,  in the determination of
$m_W/m_Z$
N--O--V employed only the direct collider data. If one also includes the
value
$\sin^2\theta_W\equiv 1-m_W^2/m_Z^2=0.2257\pm 0.0046$, obtained from
$\nu N$ scattering \cite{12}, one finds $m_W/m_Z=0.8799\pm0.0019$,
which differs from the B.A. by 1.6$\sigma$.
We also note that if we were to use  $\amzc$ instead of $\amz$ in
Eq.(\ref{sin}), we would obtain $\hat{s}_0^2=0.2337(3)$ and the
corresponding
B.A. predictions for $(g_V/g_A)_\ell$ and $m_W/m_Z$ would differ from the
experimental values  by 2.4$\sigma$.
However, in this alternative B.A. formulation, $\Gamma_\ell$, $\Gamma_h$,
and $\Gamma_Z$ are still acceptable, roughly at the 1$\sigma$ level.

Thus, we see that in the most favorable of the two B.A. approaches,
 namely the
$\amz$, $s_0^2$ scheme proposed by N--O--V, the current direct
evidence for
additional corrections involves deviations of $\approx 1.4\sigma$ in
$(g_V/g_A)_\ell$ and $\approx 1.6\sigma$ in $m_W/m_Z$ if the $\nu N$
determination is included, and is therefore not strong. However, such
 signals
are likely to be very volatile in the short run, as they depend sensitively
on the central values for $\sef$ and $m_W/m_Z$, and their corresponding
errors.

On the other hand, it is important to emphasize that there is at present
substantial indirect or inferred evidence from LEP and collider physics for
 significant corrections beyond $\arun$. Such evidence can be uncovered by
 analyzing the various observables in the framework of the complete theory,
including its complex panoply of \ewc and interlocking relations \cite{13}.
The reason is that, under such scrutiny, the SM becomes highly constrained.
In particular, the very recent analysis of Ref.\cite{11} leads to
$m_t=164{+16\atop -17}{+18\atop - 21}$ GeV (the central value is for
$m_H=300$
GeV, while the second uncertainty reflects the shifts corresponding
to $m_H=60$ GeV and 1 TeV).
As a heavy top quark decouples from $\amz$, it is clear that such
constraint
arises from the study of \ewc not contained in $\amz$.
The same observation applies to the search for signals of new physics
in quantum loop effects \cite{mar}.
 A particularly
beautiful illustration of how one can obtain strong indirect evidence
for such corrections is provided by the analysis of $m_W$. There are at
present three independent ways to determine this fundamental parameter:
i) using $m_W$ from CDF, $m_W/m_Z$ and $m_W$ from UA2, and $m_Z-m_W$
from UA1, in conjunction with the LEP value for $m_Z$, one finds $
m_W=80.23\pm0.26$ GeV
ii) employing $\sin^2\theta_W=0.2257\pm 0.0046$ from $\nu N$ scattering
\cite{12}, one has $m_W=80.24\pm 0.24$ GeV
iii) from the LEP observables, via the electroweak corrections,
one obtains the indirect determination $m_W=80.25\pm 0.10
 {+0.02\atop -0.03}$
GeV \cite{11}. The consistency of the three values and the accuracy of iii)
are
 quite remarkable. Choosing in iii) $ m_W=80.22\pm 0.10$ GeV, corresponding
to $m_H=60$ GeV (the most unfavorable option for our argument),
the weighted average of the three values becomes $m_W=80.22\pm 0.087$ GeV.
This differs from $m_W=79.95\pm 0.02$ GeV (the N--O--V B.A. prediction)
by 3$\sigma$ and from $m_W=79.82\pm 0.02$ GeV (the $\amzc$, $\hat{s}_0^2$
B.A. prediction) by 4.5$\sigma$.

This result can also be expressed in terms of $(\Delta r)_{\rm res}$
\cite{10,14}, the residual part of $\Delta r$
after extracting the effects associated with the running of $\alpha$.
Recalling  the relation \cite{4}
\be
m_W^2\left(1-\frac{m_W^2}{m_Z^2}\right)= \frac{\pi\alpha}
{\sqrt{2} G_\mu (1-\Delta r)},
\label{6}\en
we see that, given $\alpha$, $G_\mu$, and $m_Z$, a determination of $ m_W$
leads to a definite value for $\Delta r$. In particular, using
 $m_W=80.22\pm  0.087$ GeV one finds $\Delta r=0.0447\pm
0.0051$. Writing
\be
\alpha/(1-\Delta r) = \arun/(1-\dres),
\en
one has $\dres=-0.0158\pm 0.0054$ if $\arun=\amz$ (Cf. Eq.(\ref{amz}))
and $\dres=-0.0235\pm 0.0055$ if $\arun=\amzc$ (Cf. Eq.(\ref{amzc})).
We see quite clearly that $\dres$ is not zero but differs from a null
result by $\approx 2.9\sigma$ or $\approx 4.3\sigma$,
depending on how the running of $\alpha$ is parametrized.
Recalling that the natural dimensionless  coupling for \ewc
not contained in $\arun$ is $\alc/2\pi \hat{s}^2 \approx 0.54\times
10^{-2}$
($\hat{s}^2 \equiv \scr$ is the $\msbar$ parameter), we also see that the
central values for $\dres$ given above are not small.
Thus, the  current global analysis of LEP and collider physics, based on the
complete theory, points out to the existence of significant
\ewc beyond $\arun$.
There is another important theoretical advantage in using $\dres$ as a
signal for corrections ``beyond the running of $\alpha$''.
The point is that, as illustrated in Eq.(\ref{6}), $\Delta r$ is a physical
observable. This means that $\Delta r $ is renormalization--scheme
independent
and, therefore, it is not affected by the way  in which the weak
mixing angle is
 introduced. For instance, if carried out with sufficient
accuracy, theoretical calculations of $\Delta r$ should give the same
result whether one identifies the weak mixing parameter with $\scr$,
$\sef$, $s^2_0$, $\hat{s}^2_0$, or $1-m_W^2/m_Z^2$. On the other hand,
$\dres$ does depend on how $\arun$ is defined. This latter ambiguity is
unavoidable, as the analysis of the corrections not contained in $\arun$
obviously depends on the meaning of this parameter.

Accepting the results of the global analyses in the framework of the
complete theory, it is also not difficult to show that no B.A.,
whether related to $\arun$ or not, can accurately describe all the available
information. We note the current global values $\sef=0.2325\pm0.0005{+0.0001
\atop -0.0002}$, $\sin^2\theta_W=0.2257\pm0.0017{-0.0003\atop +0.0004}$
\cite{11}.
Assuming again $m_H=60$ GeV, the most unfavorable option for our
argument, we see that the two quantities differ by 3.5$\sigma$. Furthermore,
both are physical observables. Thus, it is clear that no B.A. involving a
single $s^2$ parameter can satisfactorily accomodate the values associated
with the two observables.

If the top quark is discovered and its mass measured, the indirect evidence
may become much sharper. For instance, if $m_t$ is measured to $\pm 10 $ GeV
and the corresponding central $m_W $ value remains unaltered, the
indirect LEP
determination would become approximately $m_W=80.25\pm0.06{+0.02\atop
-0.03}$ GeV. The combined result
for $m_H=60$ GeV (the most unfavorable case) would then be $m_W=
80.22\pm0.06$
GeV, which differs from the N--O--V B.A. prediction by 4.3$\sigma$.

In order to obtain stronger direct evidence one would like to improve the
 measurement of important observables such as $m_W$ and $\sef$. If, for
example, the central value of $m_W$ remains at $\approx 80.24$ GeV
(the value favored by current analyses), a measurement to $\pm 100$ MeV
would imply a difference of $2.8\sigma$ with the B.A. prediction.
It is also important to improve the direct measurement of $\sef$ from the
 on--resonance asymmetries at LEP and SLC. An increase in the central value
 or a decrease in
the error would imply a sharper discrepancy with the B.A. calculation.
This would be the case, for instance, if the central value derived from the
asymmetries  (currently 0.2321) shifts in the future towards 0.2325 (the
current global value).

In summary, in the analysis of universality \cite{2,8} there is very strong
evidence for the existence of \ewc not contained in $\arun$.
In the study of LEP and collider physics we have made a distinction between
direct and indirect or inferred evidence. In the first case, important
observables such as $m_W$, $\sef$, $\Gamma_\ell$, $\Gamma_h$, and $\Gamma_Z$
are determined directly or almost directly from experiments and then compared
 with predictions of B.A.  schemes involving $\amz$ or $\amzc$. This is the
approach followed in Ref. \cite{6,7}. In the second case precise
determinations
of fundamental parameters such as $m_W$, $\Delta r$,  $\scr$ ... are made
by analyzing the global information in terms of the complete SM, including
its
radiative corrections. We have emphasized that $\dres$ provides an important
 signal independent of the definition of the weak mixing angle.
In both cases the evidence depends sensitively on whether one employes
$\amz$ or $\amzc$ as parametrizations of the running of $\alpha$.
At LEP and collider experiments the current direct signals for corrections
beyond $\arun$ are not strong, but there is substantial indirect evidence,
which will likely become sharper when $m_t$ is determined.
We have also pointed out that there is at present considerable indirect
evidence that no B.A. can accurately describe all the available information.
Furthermore, the study of such corrections is very important in order to
constrain $m_t$ and search for signals of new physics.
In order to improve the direct evidence,
more accurate measurements of $m_W$ and
$\sef$ are called for.

\vskip 1.5cm
\noindent{\bf{Aknowledgements}}
\vskip .4cm
The author is indebted to B. King and P. Langacker for very useful
discussions and communications.
This research was supported in part by the NSF under Grant No.
PHY--9017585.

\end{document}